\documentclass[pre,aps,amsfonts,amssymb,amsmath,graphicx,showpacs,twocolumn,epsfig,amscd,floatfix]{revtex4}

\usepackage{graphicx}

\DeclareTextSymbol{\degre}{OT1}{23}

\def\bi{\begin{itemize}}

\def\ei{\end{itemize}}
\def\be{\begin{equation}}
\def\ee{\end{equation}}
\def\bea{\begin{eqnarray}}
\def\eea{\end{eqnarray}}
\def\f{F_{min}}
\def\F{F_{max}}
\def\fs{\bar{F}}
\def\tstick{t_{stick}}

\begin{document}

\author{G. Ovarlez, E. Cl\'{e}ment}
\affiliation{Laboratoire des Milieux D\'{e}sordonn\'{e}s
et\\H\'{e}t\'{e}rog\`{e}nes - UMR 7603\\Universit\'{e} Pierre et
Marie Curie - Bo\^{\i}te 86, 4, Place Jussieu, F-75252\ Paris}
\title{Slow dynamics and aging of a confined granular flow}
\date{}

\begin{abstract}
We present experimental results on slow flow properties of a
granular assembly confined in a vertical column and driven upwards
at a constant velocity V. For monodisperse assemblies this study
evidences at low velocities ($1\!<\!V\!<\!100\ \mu m/s$) a
stiffening behaviour i.e. the stress necessary to obtain a steady
sate velocity increases roughly logarithmically with velocity. On
the other hand, at very low driving velocity ($V\!<\!1\ \mu m/s$),
we evidence a discontinuous and hysteretic transition to a
stick-slip regime characterized by a strong divergence of the
maximal blockage force when the velocity goes to zero. We show
that all this phenomenology is strongly influenced by surrounding
humidity. We also present a tentative to establish a link between
the granular rheology and the solid friction forces between the
wall and the grains. We base our discussions on a simple
theoretical model and independent grain/wall tribology
measurements. We also use finite elements numerical simulations to
confront experimental results to isotropic elasticity. A second
system made of polydisperse assemblies of glass beads is
investigated. We emphasize the onset of a new dynamical behavior,
i.e. the large distribution of blockage forces evidenced in the
stick-slip regime.
\end{abstract}
 \maketitle

\section{Introduction}

Granular flows are presently at the focus of many attentions
\cite{PDM}. The classical model for granular media stability was
proposed at the end of the XVIIIth century by Ch. A. de Coulomb
who revealed a strong analogy between the failure properties of a
granular assembly and the phenomenology of solid on solid
friction. Modern developments have elaborated sophisticated
empirical approaches around this fundamental idea \cite{Wood} but
so far there is no deep physical understanding nor rigorous
derivation describing the passage from the granular level
description to a set of evolution equations involving macroscopic
quantities such as stress, strain or packing fraction.
Furthermore, there is an additional difficulty to apprehend
complex behaviors such as aging under stress or to account for the
influence of external parameters such as surrounding humidity
which effects are often noticed in practice. Note that similar
questions are still under active consideration in the field of
tribology \cite{Berthoud99,Crassous99,Riedo02}, but in the case of
granular assemblies, a supplementary difficulty lies in the
fragile character of granular structures which can be easily
modified under the action of external constraints.

Dynamical behavior of slowly driven granular materials was
investigated by different groups both in compression and/or in
shearing experiments \cite{Gollub97,Horwarth96,Albert99,Lubert}.
Here we present an experimental situation of the same type, but in
a quite different geometry. We investigate the rheology of a
granular assembly confined in a cylindrical column and pushed
vertically from the bottom. The resistance to vertical motion as
well as the blocking/unblocking transitions reveals a
phenomenology possibly shared by many confined granular
assemblies. Note that this column configuration may help to
understand several practical situations like pipe flows
\cite{Hulin}, compaction under stress or dense granular paste
extrusion. A previous investigation of the same display was made
in 2D \cite{LaKolb99} and also in 3D \cite{Ovarlez01} as a
preliminary report, and more recently see refs
\cite{Bertho03,Landry}. Those contributions have shown a rich
phenomenology partly sorted by the solid friction properties of
the grains and the boundaries. The importance of surrounding
humidity was also evidenced \cite{Ovarlez01} as it would strongly
influence the rheology of the granular column. In this report, we
push further the investigation as we change the column material,
the beads characteristics and size dispersity, under various
humidity conditions. We also propose to compare our data with the
outcome of a simple numerical model of isotropic elasticity.

In the classical situation of newtonian fluid pushed by a piston
(typically in a syringe), one would obtain a relation between
pressure and flow rate, which is characterized by a fluid
constitutive parameter: the viscosity. For an isotropic elastic
medium the resistance to pushing would depend on the material
Poisson's ratio and wall frictional properties
\cite{Ovarlez03,OvarlezSimuls} (we will detail this question
further in the text). For a granular material, the situation is a
priori more complex since the piling structure can be modified so
as to adapt to the external constraints. Friction at the walls may
also create internal granular recirculation flows \cite{LaKolb99}.
Therefore it is an important but difficult matter to clarify the
rheology of this system by sorting the respective influence of
grain-boundary friction versus bulk structural changes. Along this
line we started with a simple situation of weakly frictional steel
beads with frictional boundaries. A second situation is studied
where the granular material is made of rugous polydisperse glass
beads.

\section{Display}

\begin{figure}
\includegraphics[width=5cm,angle=-90]{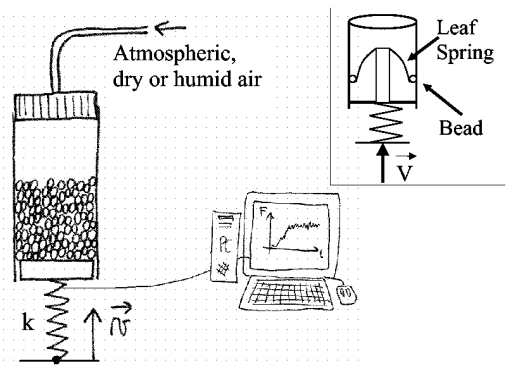}
\caption{Sketch of the experimental display. Inset: sketch of the
slider.} \label{Fig1}
\end{figure}

Most experiments are performed on dry, non cohesive and
monodisperse steel beads of diameter $d=1.58$ mm piled into a
vertical duralumin or brass cylinder of diameter $D=36$ mm. We
also perform a series of experiments on a miscellany of glass
beads of three diameters (1.5 mm, 2 mm and 3 mm) with equal volume
of each kind, in a PMMA cylinder. The column is closed at the
bottom by a movable brass piston avoiding contact with the column
(diameter mismatch is $0.5$ mm). A force probe of stiffness
$k=40000$ $\mbox{N.m}^{-1}$ is located under the piston and is
pushed at a constant driving velocity $V$ (between $5$
$\mbox{nm.s}^{-1}$ and $100$ $\mu\mbox{m.s}^{-1}$) via a stepping
motor (see Fig. \ref{Fig1}). The resistance force $F$ encountered
by the piston is measured as a function of time. We also monitor
the relative humidity ($\chi$) and the surrounding temperature. We
work in the range $35\%<\chi<75\%$, as well as in dry air
($\chi<3\%$) and humid air ($\chi=90\%$). Actually, except for the
dry and humid situation, we do not regulate this last parameter
($\chi$) but we record its values close to the experimental
set-up. We obtain dry air by having a weak air flux ($250\
\mbox{ml.min}^{-1}$) flow into a cylinder filled with silica. We
obtain humid air by making a weak air flux ($250\
\mbox{ml.min}^{-1}$) bubble through water. Then, for the duration
of an experiment, weak fluxes of humid or dry air are set to flow
through the column from top to bottom, maintaining a constant and
homogeneous level of humidity. Note that a stationary relative
humidity is reached in a few minutes. We start our experiments
when this stationary level is attained. We verified that the weak
air flux does not perturb the system: we notice no evolution of
the force when the flux is stopped as long as relative humidity
remains unchanged. Temperature is kept at $(20\pm1)$\degre C. We
actually find no correlation between the force fluctuations and
the temperature variations in this range.

\begin{figure}
\includegraphics[width=7cm]{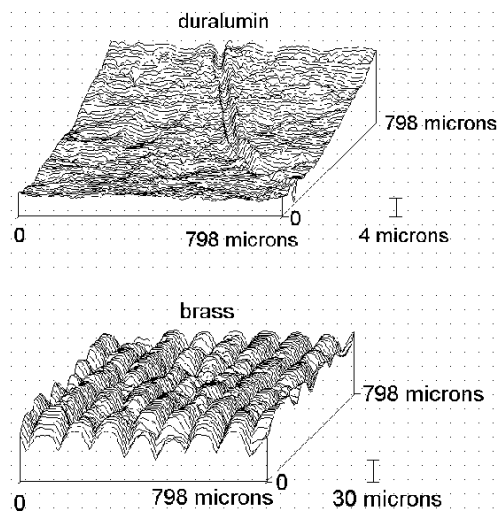}
\caption{Topographies of internal surfaces of duralumin and brass
column used in our experiments. The surfaces were scanned on $800\
\mu m$ by $800\ \mu m$, with $2\ \mu m$ steps; vertical resolution
is 100 nm. The horizontal axis on the graphes is the symmetry axis
of the columns.} \label{Fig2}
\end{figure}

Note that the brass and duralumin cylinders have very different
surface properties (Fig. \ref{Fig2}). The duralumin cylinder is
rough cast, and its roughness is $400\ $nm. The brass cylinder was
machine-turned, and has mean roughness $7\ \mu$m with undulations
of wavelength $100\ \mu$m and depth $25\ \mu$m.

In order to compare precisely rheological properties of granular
assemblies to solid friction properties, we built a special device
(called ''the slider'', see Fig. \ref{Fig1} inset) designed to
study the tribology of bead/wall contact. This device is set to
apply a constant normal load ($F_{N}=2$ $N$) on three steel beads
sliding vertically on the cylinder's wall. Then, the dynamical
evolution of the resistance force encountered by the piston
pushing a granular material can be compared to the slider's
friction resistance driven in the same conditions.

Two granular systems will be studied. First a model assembly of monodisperse
steel spheres and second a polydisperse assembly of rough glass beads. The
largest part of this report is devoted to the monodisperse assembly.

\section{Monodisperse steel beads}

\begin{figure}
\includegraphics[width=7cm]{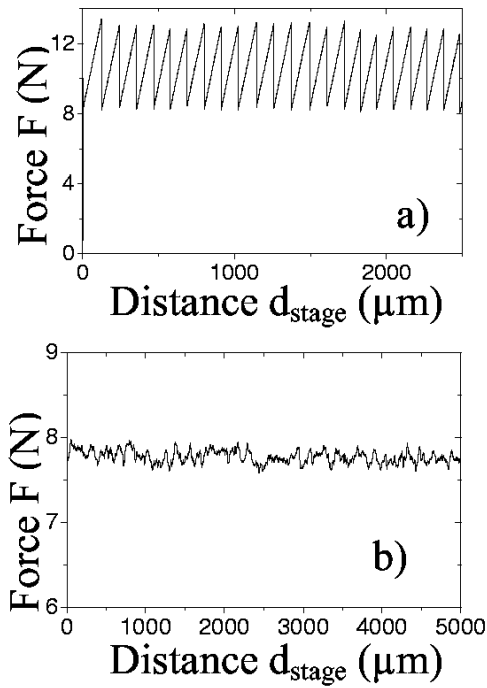}
\caption{Resistance force to pushing of a height $H=2.15D$ (380 g)
of steel beads in the duralumin cylinder vs. displacement
$d_{stage}$ of the translation stage. a: in the stick-slip regime
($V=30\ \mbox{nm.s}^{-1}$). b: in the steady-sliding regime
($V=100\ \mu\mbox{m.s}^{-1}$).} \label{Fig3}
\end{figure}

We now report on the simplest situation i.e. monodisperse low
frictional steel beads in a duralumin or brass cylinder. We
observe two distinct dynamical stationary regimes (Fig.
\ref{Fig3}): for high driving velocities, the motion is
characterized by a steady-sliding and a constant pushing force;
for low velocities, the system undergoes a dynamic instability
characterized by a stick-slip motion.

\begin{figure}
\includegraphics[width=7cm]{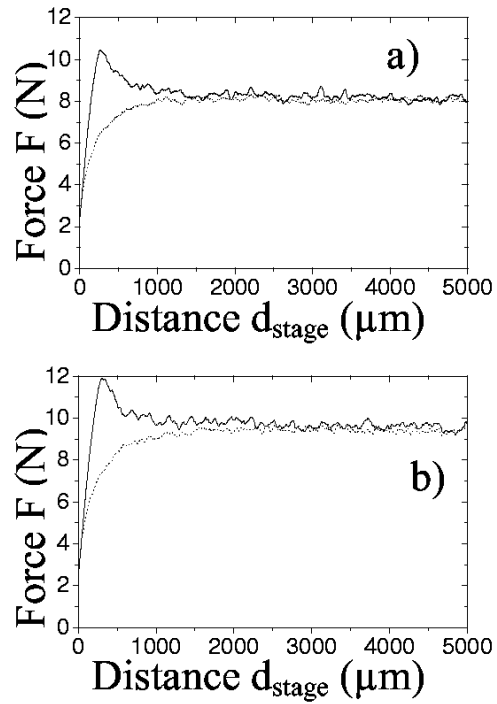}
\caption{Evolution of the resistance force in the steady-sliding
regime for two different initial packings: a dense one (65\%) and
a loose one (59\%); the full line is for the initially dense
packing, the dotted line is for the initially loose packing. a)
$V=20\ \mu\mbox{m.s}^{-1}$; b) $V=100\ \mu\mbox{m.s}^{-1}$.}
\label{Fig4}
\end{figure}

We first notice that as far as averaged packing fraction and
pushing force are concerned, whatever the initial state of
preparation is, the same stationary regime is reached (Fig.
\ref{Fig4}). When a dense packing
($\overset{\_}{\nu}=65.0\pm0.5\%$) is prepared by rain-filling,
the force in the steady sliding regime increases up to a maximum,
then decreases slowly for a piston displacement of about 3 mm. At
this point where a stationary value $\bar{F}$ of the force is
attained, the average packing fraction $\overset{\_}{\nu}$ is
$62.5\pm0.5\%$. In the case of an initially loose packing
($\overset{\_}{\nu}=59.0\pm0.5\%$) prepared by using an inner
cylinder slowly removed after filling, the force in the steady
sliding regime increases monotonically for a 3 mm displacement
before the same stationary value $\bar{F}$ of force is reached.
This state is also characterized by a $62.5\%$ packing fraction.
For all preparations, we always get a stationary regime
characterized by the same pushing force $\bar{F}$, which depends
on driving velocity $V$, relative humidity $\chi$ and packing's
height $H$, as well as the same packing fraction
($\overset{\_}{\nu}=62.5\pm0.5\%$), independent of $V$, $\chi$ and
$H$. The stick-slip regime displays a similar phenomenology: the
piling reaches a stationary packing fraction of
$\overset{\_}{\nu}=62.5\pm0.5\%$; in the transitory regime, the
pushing force for dense packing displays a maximum and for a loose
packing shows a monotonous increase. In section \ref{III.A.1} we
propose an explanation for the phenomenology of the transitory
regimes.

\vspace{0.5cm} In the following, we present the phenomenology
observed in the steady-sliding and the stick-slip regime, when the
stationary state is attained. For each regime, we first focus on
the mechanical properties and propose a model accounting for
friction at the walls and bulk properties; we then study and
analyse the rheological properties. In a third section, we point
out ambiguity of dependence on relative humidity. We then analyse
our results in a standard solid on solid friction framework (the
Dieterich-Ruina model). Finally, we study the transition from
steady-sliding to stick-slip.

\subsection{Steady-sliding}

\begin{figure}
\includegraphics[width=7cm]{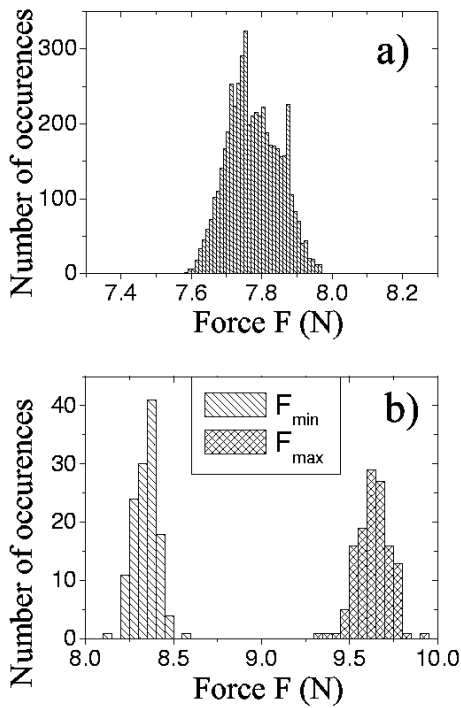}
\caption{a: Distribution of the resistance force value for every
$\mu$m during 5 mm of sliding in the steady-sliding regime of Fig.
\ref{Fig3} at $V=100\ \mu\mbox{m.s}^{-1}$. b: Distribution of the
maximum and minimum resistance forces $\F$ and $\f$ in the
stick-slip regime of Fig. \ref{Fig3} at $V=30\ \mbox{nm.s}^{-1}$
during 5 mm of sliding.} \label{Fig5}
\end{figure}

For a given height of beads in the column, at given driving
velocity $V$ and relative humidity $\chi$, the force in the steady
sliding regime is constant at about $2\%$ (Fig. \ref{Fig5}a), and
its distribution around mean value $\bar{F}$ is nearly gaussian.

Next, we study the mean pushing force $\bar{F}$ behavior as a function of the
packing height, the driving velocity and the relative humidity.

\subsubsection{Mechanical properties}\label{III.A.1}

\begin{figure}
\includegraphics[width=5cm,angle=-90]{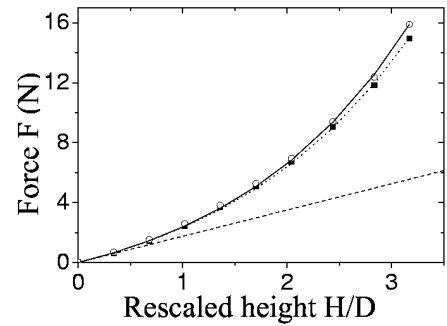}
\caption{Mean resistance force in the steady sliding regime as a
function of the height $H$ of the packing scaled by the column
diameter $D$, for steel beads in the duralumin column, for $V=16\
\mu\mbox{m.s}^{-1}$ (filled squares) and $V=100\
\mu\mbox{m.s}^{-1}$ (open circles). The line and the dotted line
are fits by eq. (\ref{1}). The dashed line is the hydrostatic
curve.} \label{Fig6}
\end{figure}

For a vertically pushed granular assembly, the driving force
exerted by the piston is screened due to friction with the walls.
To evaluate this effect, the mean resistance force $\bar{F}$ in
the steady-sliding regime is measured as a function of the packing
height (see Fig. \ref{Fig6}).

The resistance force $\bar{F}$ increases very rapidly with the
packing's height $H$. Following the standard Janssen screening
picture, this strong resistance to motion is due to the leaning of
the granular material on the walls created by the horizontal
redirection of vertical stress in association with solid friction
at the sidewalls. It means that we may relate horizontal and
vertical stresses averaged on a slice at height $z$ by an
effective relation: \be\sigma_{zz}(z)=K\sigma_{rr}(z)\ee where $K$
is called Janssen's redirection constant. At the walls, we suppose
a sliding of the granular material at a velocity $V_{0}$ (the
driving velocity); the shearing stress is then
\be\sigma_{rz}(z)=\epsilon\mu_{d}(V_{0})\sigma_{rr}(z)\ee where
$\mu _{d}(V_{0})$ is the dynamic coefficient of friction between
the beads and the cylinder's wall at a velocity $V_{0}$. The
constant $\epsilon=\pm1$ is introduced in order to differentiate
between pushing and pulling experiments. When $\epsilon=+1,$ the
granular material is moving upwards and friction is fully
mobilized downwards (our pushing experiment) and when
$\epsilon=-1,$ the granular is moving downwards and friction is
fully mobilized upwards.

The force $\bar{F}_{\varepsilon}$ exerted by the grains on the
piston can be derived from equilibrium equations for all slices,
thus we obtain: \begin{equation} \bar{F}_{\epsilon}=\varrho
g\lambda\pi R^{2}\times\epsilon(\exp(\epsilon
\frac{H}{\lambda})-1) \label{1}
\end{equation}
where $\varrho$ is the mass density of the granular material, $R$ is the
cylinder radius and $g$ the acceleration of gravity. The length $\lambda
=R/2K\mu_{d}(V_{0})$ is the effective screening length.

It is easily seen from (\ref{1}) that when $\epsilon=+1$, any
slight change in $\mu$ or $K$ is exponentially amplified with a
drastic influence on the pushing force $\bar{F}$. In the steady
state regime, the experimental data obtained for a given pushing
velocity $V$ can be fitted by relation (\ref{1}) by adjusting only
one parameter i.e. \be p_{+1}=K\times\mu_{d}(V_{0})\ee where
$\mu_{d}(V_{0})$ is the dynamic coefficient of friction at
velocity $V_{0} $.\newline We see on Fig. \ref{Fig6} that the data
are well fitted by eq. (\ref{1}), and, for a relative humidity
$\chi=45\%$, for steel beads in the duralumin cylinder, we obtain
$p_{+1}=0.140\pm0.001$ at $V_{up}=16\ \mu \mbox{m.s}^{-1}$ and
$p_{+1}=0.146\pm0.001$ at $V_{up}=100\ \mu\mbox{m.s} ^{-1}$.

We are now able to propose an interpretation of the features
observed for different initial packing fractions within Janssen's
framework. Since Janssen's coefficient $K$ is an increasing
function of the packing fraction, as it was shown previously
\cite{Vanel99,Ovarlez03}, the resistance force of an initially
dense packing displays a maximum, as a dense packing leans more
efficiently on walls than a loose one; the pushing force then
decreases while the packing loosens (i.e. while $K$ decreases). On
the other hand, for an initially loose packing the resisting force
in the transitory regime will increase while the packing densifies
(i.e. while $K$ increases).\newline For an initially dense
packing, we extract a parameter $p_{+1}^{max}$ from a fit of the
pushing force maximum value in the transitory regime with formula
(\ref{1}): we obtain $p_{+1}^{max}\approx0.18$ at $V_{up}=16\
\mu\mbox {m.s}^{-1}$ and $p_{+1}^{max}\approx0.18$ at $V_{up}=100\
\mu\mbox{m.s}^{-1}$; as this experiment was not repeated
sufficiently, we do not have uncertainties on these values.
Nevertheless, we notice that the $p_{+1}^{max}$ value is roughly
25\% higher than the $p_{+1}$ stationary value, whatever the
velocity is. These results are consistent with a $K$ dependence on
compacity $\overset{\_}{\nu}$ derived in \cite{Ovarlez03} for an
assembly of monodisperse glass beads, i.e. $\Delta
K/K\approx5\Delta\overset{\_}{\nu }/\overset{\_}{\nu}$; for
initial compacity $65\%$ and steady-state compacity $62.5\%$, this
empirical formula actually leads to $\Delta K/K\approx0.2$.
Therefore, in the Janssen framework, we are led to attribute the
difference between maximum force and stationary force to a
difference in stress redirection (i.e. in $K$) due to a difference
in compacity. The force history in the transitory regime would
then just reflect the compacity history.

\begin{figure}
\includegraphics[width=5cm,angle=-90]{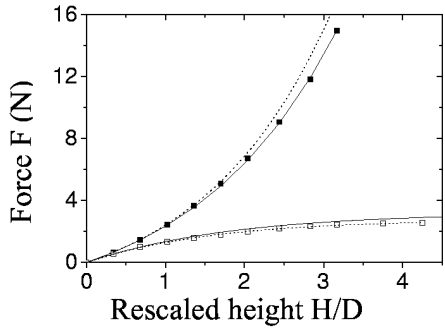}
\caption{Resistance force to pushing (filled squares) and to
pulling (open squares) in the steady-sliding regime at $V=16\
\mu\mbox{m.s}^{-1}$ as a function of the height $H$ of the packing
scaled by the column diameter $D$, for steel beads in the
duralumin column. The lines are the fit with eq. (\ref{1}) of the
resistance force to pushing and its prediction for the pulling
situation; the dotted lines are the fit with eq. (\ref{1}) of the
resistance force to pulling and its prediction for the pushing
situation.} \label{Fig7}
\end{figure}

As a check of consistency, we perform the following dynamical
experiment. First, the granular column is pushed upwards in order
to mobilize the friction forces downwards and far enough to reach
the steady state compacity. Starting from this situation, the
friction forces at the walls are reversed by moving the piston
downwards at a constant velocity $V_{down}=16\
\mu\mbox{m.s}^{-1}$, until a stationary regime is attained. Note
that this stationary regime is characterized by the same compacity
$\overset{\_}{\nu}\approx62.5\%$ as in the pushing situation.
Following relation (\ref{1}), this procedure would imply a change
of $\epsilon$ from $1$ to $-1$, and consequently, the dynamical
force on the piston should decrease from $\bar{F}_{+1}$ to
$\bar{F}_{-1}$. In Fig. \ref{Fig7} the pushing force
$\bar{F}_{-1}$ is measured for different packing heights $H$. The
fit of experimental results with eq. (\ref{1}) gives $p_{-1} (16\
\mu\mbox{m.s}^{-1})=0.156\pm0.002$ which is $10\%$ larger than
$p_{+1}(16\ \mu\mbox{m.s}^{-1})$. This difference, though small,
can be observed out of uncertainties, and is systematic. It cannot
be due to a slight change in compacity $\overset{\_}{\nu}$ as from
relation $\Delta
K/K\approx5\Delta\overset{\_}{\nu}/\overset{\_}{\nu}$, we would
expect a 2\% variation in compacity between the pushing and the
pulling experiment, which would be observed; we actually measured
$\Delta\overset{\_}{\nu}/\overset{\_}{\nu}=0\pm1\%$. According to
Janssen's picture, this would imply that vertical stress
redirection is more efficient in the downward pulling situation.
We believe this is a clear evidence of a granular structuration
effects but its also shows that this effect is not dominant: it
affects only $10\%$ of the average mechanical parameter $K$.

We have shown in another report that in a Janssen experiment, an
isotropic homogeneous elastic material can also be characterized
by stress redirection properties \cite{Ovarlez03}. For a granular
column at high depth an effective relation:
\be\sigma_{zz}=K_{el}\sigma_{rr}\ee is obtained with \be
K_{el}=\nu _{p}/(1-\nu_{p})\ee where $\nu_{p}$ is the material's
Poisson coefficient. This actually leads to a curve similar to
Janssen's saturation curve as long as friction at the walls is
small (typically less than $0.5$).\\
In order to get the isotropic homogeneous elasticity prediction
for the pushing experiment, we perform a series of numerical
simulations using Finite Element Method \cite{Castem}. The column
is modelled as an isotropic elastic medium. We vary the friction
at the walls $\mu_{d}$, the Young modulus $E$ and the Poisson
coefficient $\nu_{p}$. We impose a rigid, either perfectly stick
or perfectly slip bottom. We find no appreciable difference
between these two previous cases. The condition
$\sigma_{rz}=-\mu_{d}\sigma_{rr}$ is imposed everywhere at the
walls (for the pulling situation, we impose $\sigma_{rz}=+\mu_{d}
\sigma_{rr}$). The cylinder is modelled as a duralumin elastic
medium. As long as the Young modulus $E$ of the elastic medium is
less than $500$ MPa, which is usually the case for granular media,
we find no dependence of the results on $E$. We verified that in
all the simulations we performed, there is no traction in the
elastic medium, so that this could be a fair modelling for a
granular material.

\begin{figure}
\includegraphics[width=5cm,angle=-90]{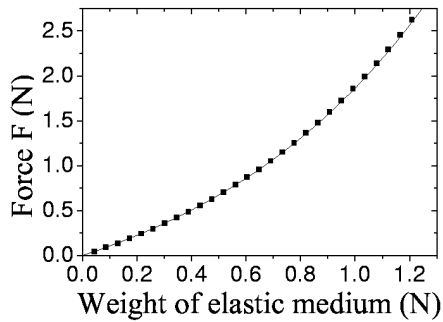}
\caption{Comparison of the resistance force to pushing simulated
for a homogeneous isotropic elastic medium (squares) of Poisson
coefficient $\nu_p=0.45$ and Young modulus $E=100$ MPa in a
duralumin cylinder, with coefficient of friction $\mu=0.2$ at the
walls, to the curve obtained with eq. (\ref{1}) with Janssen
coefficient $K=K_{el}=\nu_p/(1-\nu_p)=0.82$ (full line).}
\label{Fig8}
\end{figure}

We find no appreciable difference between the elastic prediction
(Fig. \ref{Fig8}) and the curve given by eq. (\ref{1}) with
$K=K_{el}$. Therefore, regarding the dependence of the stationary
state force $\bar{F}$ on the height of beads, our system cannot be
distinguished from an elastic medium.

Note that finite element simulations show that the presence of a
rigid bottom implies that the effective Janssen's parameter
$K_{eff}$ extracted from Janssen's scaling for the pulling
situation is higher than $K_{el}$ \cite{Ovarlez03}, whereas for
the pushing $K_{eff}\approx K_{el}$ (as can be seen on Fig.
\ref{Fig8}: the fit of the elastic curve with $K=K_{el}$ is good).
Details on this numerical work will be given elsewhere
\cite{OvarlezSimuls}. Actually, if we adjust the elastic
predictions for pushing and pulling experiments with an elastic
material of Poisson coefficient $\nu_{p}=0.45$, eq. (\ref{1})
yields a Janssen's constant $K_{eff}$ for the pushing which is
about $3\%$ lower than $K_{eff}$ for the pulling. This is
qualitatively (though not quantitatively) in agreement with the
experimental results. Then isotropic elasticity can be a good
framework only if we neglect the existence of bulk structuration
effects inducing differences in the effective Poisson coefficient
of the material between the pulling and the pushing. Note that in
this case, an isotropic modelling of the granular material is
somehow questionable.

\vspace{0.2cm} In a previous study \cite{Vanel99}, it was found
that the Janssen picture has a general tendency to slightly
underestimate the stress below a granular column for a homogeneous
packing of glass beads. We have showed in another report that it
is no more true as the friction at walls is very well controlled.
Therefore, this model, though elementary, seems a fair base for
analysis and provides an analytical expression from which
constitutive rheological parameters can be extracted. A central
question is still that the fitting parameters $p=K\times\mu_{d}$
extracted from the model does not allow to distinguish between
$\mu_{d}$ and $K$ separately.

In the following, we will show experiments which aim at sorting
out the relative contributions of wall-bead interactions (i.e.
$\mu_{d}$) and bulk properties (i.e. $K$) which have an influence
on the rheological properties when velocity and relative humidity
are changed.

\subsubsection{Rheological properties}

\begin{figure}
\includegraphics[width=7cm]{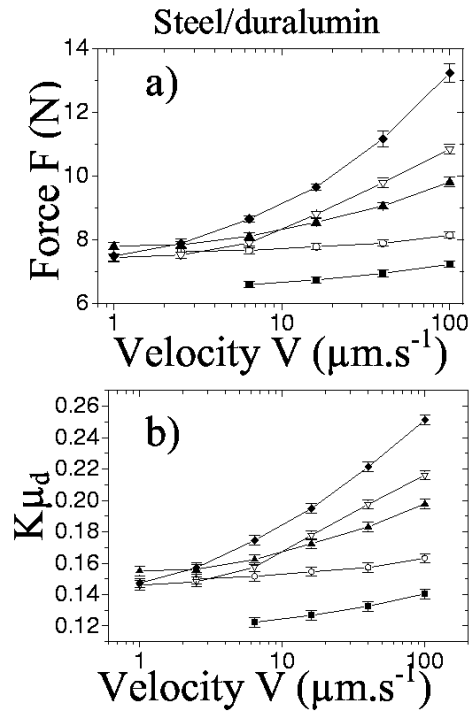}
\caption{a) Mean force $\fs$ in the steady sliding regime for
height $H=2.15D$ (380 g) of steel beads in the duralumin cylinder,
for various relative humidities $\chi$: $72\%$ (diamonds), $66\%$
(open inverted triangles), $53\%$ (triangles), $40\%$ (open
circles), and $<3\%$ (squares).  b) Coefficient $K\mu_d$ extracted
from a fit of $\fs$ with formula (\ref{1}).} \label{Fig9}
\end{figure}

\begin{figure}
\includegraphics[width=7cm]{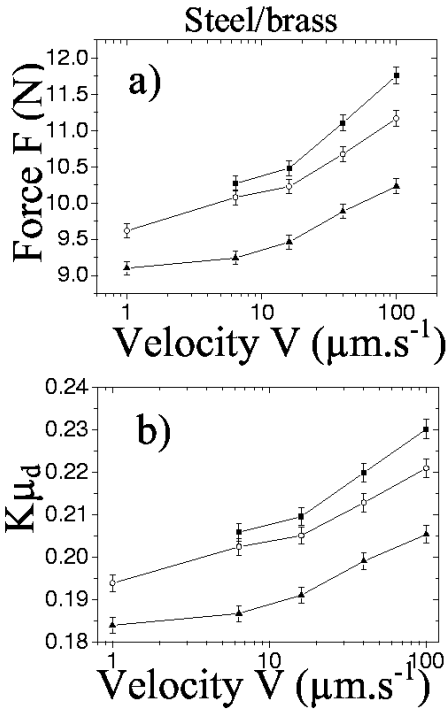}
\caption{a) Mean force $\fs$ in the steady sliding regime for
height $H=2.15D$ (380 g) of steel beads in the brass cylinder, for
various relative humidities $\chi$: $90\%$ (squares), $53\%$ (open
circles), and $<3\%$ (triangles). b) Coefficient $K\mu_d$
extracted from a fit of $\fs$ with formula (\ref{1}).}
\label{Fig10}
\end{figure}

The mean resistance force $\bar{F}$ in the steady-sliding regime
increases strongly when the velocity and the relative humidity are
increased. In the duralumin column (Fig. \ref{Fig9}), the
resistance force $\bar{F}$ at velocity $V=100\ \mu\mbox{m.s}^{-1}$
is $60\%$ higher for relative humidity $\chi=73\%$ than for
$\chi=40\%$; near the transition to the stick-slip regime ($V=1\
\mu\mbox{m.s}^{-1}$), dependence of $\bar{F}$ on $\chi$ is much
less important, and there is actually no difference in $\bar{F}$
values for $\chi$ varying between $40\%$ and $73\%$. Another
important feature is that the increase of $\bar{F}$ with velocity
is sharper when $\chi$ is higher. In the brass column (Fig.
\ref{Fig10}), the mean resistance force $\bar{F}$ also increases
with relative humidity $\chi$, but now the variation coefficient
with velocity (the slope on Fig. \ref{Fig10}) does not seem to
depend on $\chi$. Moreover, for a velocity $V=100\
\mu\mbox{m.s}^{-1}$, the $\bar{F}$ increase is only $15\%$ from
dry to humid ($\chi=90\%$) atmosphere.

The fit of $\bar{F}$ with Janssen's formula adapted to the pushing
case of eq. (\ref{1}) gives parameter $K\mu_{d}(V)$ for different
relative humidities $\chi$. The fundamental differences of
behavior in the steady sliding regime between the duralumin column
and the brass column for a same granular material, at same
density, suggest that the variations in $\chi$ and velocity have
an influence mainly on the coefficient of friction $\mu_{d}$ at
the walls, and little on mechanical properties of the granular
material (i.e. $K$). Therefore, in the following analysis, we will
consider $K$ as a constant at first order of approximation.

\subsubsection*{\textit{Steel/duralumin}}

Fig. \ref{Fig9} shows that dynamical parameter $K\mu_{d}$ is
globally less important in a dry atmosphere than in the ambient
atmosphere. In ambient atmosphere ($40\%<\chi<73\%$), a change in
$\chi$ seems to change only the variation coefficient of
$K\mu_{d}$ with velocity: i.e. the increase of $K\mu_{d}$ with
velocity is sharper when $\chi$ is higher. This phenomenon could
be interpreted as a viscous contribution of the water condensed at
the contacts. Note that this phenomenology contrasts with the
observations of Riedo et al. \cite{Riedo02} who find, for solid on
solid nanoscopic sliding friction measurements, that in all the
systems they study, of various wettability, the effect of a
humidity increase is to add a negative value to the coefficient of
friction force dependence on logarithm of velocity. Table
\ref{tablestrengtheningdural} shows the results of a rough
logarithmic fit of data with
$\mu_{d}(V)=\mu_{0}+(a\!-\!b)\ln(V/V_{0})$. We use the parameter
$(a\!-\!b)$ as a standard reference to the Dieterich-Ruina model
\cite{Dieterich79,Ruina}.

\begin{table}
\begin{center}
\begin{tabular}
[c]{|c|c|}\hline
relative humidity $\chi$ & $K(a\!-\!b)$\\\hline
$<3\%$ & $0.0065\pm0.0006$\\
$40\%$ & $0.004\pm0.0005$\\
$53\%$ & $0.013\pm0.001$\\
$66\%$ & $0.021\pm0.0005$\\
$72\%$ & $0.028\pm0.0015$\\\hline
\end{tabular}
\end{center}
\caption{Parameter $K(a\!-\!b)$ extracted from a fit of $K\mu_{d}$
with formula $\mu_{d}(V)= \mu_{0}+(a\!-\!b)\ln(V/V_{0})$ for
different relative humidities $\chi$ in the duralumin column.}
\label{tablestrengtheningdural}
\end{table}

We observe that the increase rate $K(a-b)$ of $K\mu_{d}$ with the
logarithm of velocity is multiplied by $7$ when $\chi$ increases
from $40\%$ to $72\%$.

\begin{figure}
\includegraphics[width=5cm,angle=-90]{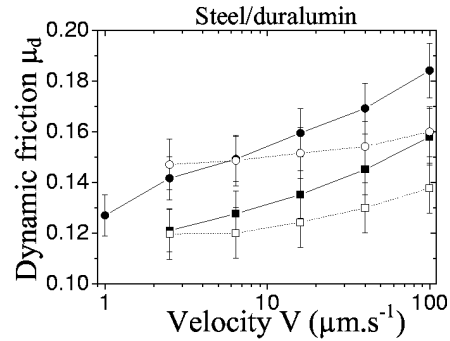}
\caption{Dynamical coefficient of friction $\mu_d$ extracted from
a fit of $\fs$ with Janssen's formula adapted to pushing (open
symbols) for a height $H=2.15D$ (380 g) of steel beads in the
duralumin cylinder, and $\mu_d$ measured in a solid on solid
experiment (filled symbols), for two relative humidities $\chi$:
$40\%$ (circles) and $<3\%$ (squares).} \label{Fig11}
\end{figure}

In order to compare these results with those obtained in a solid
on solid friction experiment, we need to evaluate Janssen's
parameter $K$. We extract $K$ from a classical Janssen experiment
\cite{Ovarlez03} and we obtain: $K\mu_{s} =0.184\pm0.002$ for a
mean aging time of $40$ seconds. From the measurement of static
friction, using the sliding angle of a three bead tripod, we
extract $K=1.02\pm0.07$.\newline We suppose $K$ is constant and is
not affected by any variation in relative humidity nor velocity.
The dynamical coefficient of friction obtained in a solid on solid
friction experiment with the slider (Fig. \ref{Fig1} inset) is
then compared on Fig. \ref{Fig11} with the one extracted from a
fit of the resistance force with formula (\ref{1}). Note that the
uncertainty on $\mu_{d}$ extracted from formula (\ref{1}) is
mainly systematic (coming from uncertainties on $K$, which come
themselves from uncertainty on static friction measurements). We
observe that the dynamical coefficients of friction measured in a
solid on solid experiment increase slightly more strongly with
velocity than the one extracted from the resistance force of the
granular material. But importantly, the increase of $\mu_{d}$ with
relative humidity $\chi$ is the same in both cases.\\ In this
analysis we assumed a constant value for Janssen's coefficient
$K$. We can also imagine a slight decrease of $K$ with increasing
velocity so that the solid on solid $\mu_{d}$ would match the one
extracted from the pushing experiment. However, we will remain
prudent as the mean pressure contact on beads in the solid on
solid experiment is much higher than in the granular column, which
may cause quantitative differences between friction properties.
More precisely, at the bottom of the column, for a pushing force
$\bar{F}=10\ $N, the mean contact force per bead is about $20\
$mN; a Hertz contact would give mean contact pressure $\bar{p}
\simeq300\ $MPa. For a contact force per bead of $2\ $N (the
slider case), a Hertz contact would give $\bar{p}\simeq1300\ $MPa.

\subsubsection*{\textit{Steel/brass}}

In the brass column, we observe the same phenomenology as in the
duralumin one, i.e. an increase of $K\mu_{d}$ with velocity and
relative humidity $\chi $. The effect of a change in $\chi$ is yet
less important than in the duralumin column, as the coefficient
$K\mu_{d}$ increases only by $15\%$ at velocity $V=100\
\mu\mbox{m.s}^{-1}$ when $\chi$ increases from $0\%$ to $90\%$. A
major difference with the duralumin column is that in the brass
column, a variation in $\chi$ seems to induce a variation of
friction but hardly affects the coefficient of variation with
velocity (the slope). Table \ref{tablestrengtheninglaiton} shows
the results of a logarithmic fit of the data with the function:
$\mu_{d}(V)=\mu_{0}+(a\!-\!b)\ln(V/V_{0})$. We observe indeed that
the slope $K(a-b)$ of $K\mu_{d}$ with the logarithm of velocity
does not practically vary with relative humidity $\chi$, and is
much weaker than in the duralumin column.

\begin{table}
\begin{center}
\begin{tabular}
[c]{|c|c|}\hline
relative humidity $\chi$ & $K(a-b)$\\\hline
$<3\%$ & $0.007\pm0.001$\\
$53\%$ & $0.007\pm0.001$\\
$90\%$ & $0.0085\pm0.0015$\\\hline
\end{tabular}
\end{center}
\caption{Parameter $K(a\!-\!b)$ extracted from a fit of $K\mu_{d}$
with formula $\mu_{d}(V)= \mu_{0}+(a\!-\!b)\ln(V/V_{0})$ for
different relative humidities $\chi$ in the brass column.}
\label{tablestrengtheninglaiton}
\end{table}

\vspace{0.3cm}

The comparison with the slider, and the differences between the
duralumin and bras columns, suggest that the rheological
properties in the steady-sliding regime are dominated by the
friction properties at the walls. We now investigate the
stick-slip regime.

\subsection{Stick-slip regime}

\begin{figure}
\includegraphics[width=7cm]{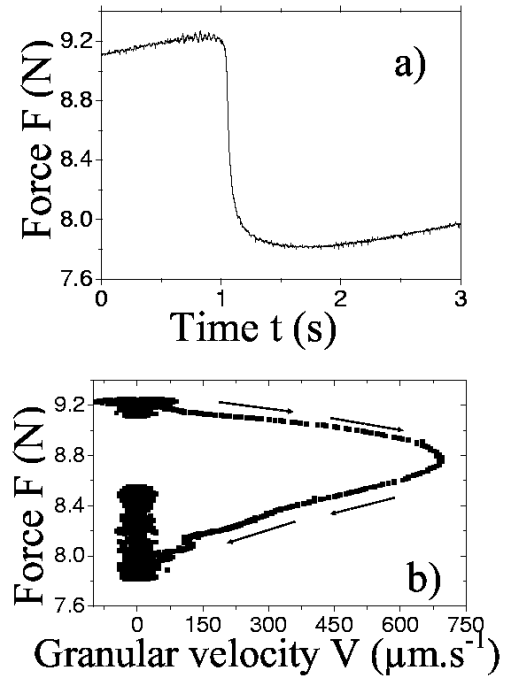}
\caption{a: Force vs. time for a slip event in the stick-slip
regime at $V=1\ \mu\mbox{m.s}^{-1}$ for a height $H= 2.15D$ (380
g) of steel beads in the duralumin column. b: Force vs. velocity
of the granular material for the slip event of Fig. a}
\label{Fig12}
\end{figure}

In the stick-slip regime, for a given height of beads in the
column and at given driving velocity $V$ and relative humidity
$\chi$, the maximum force before a slip and the minimum force at
the end of a slip are constant within $2\%$ (Fig. \ref{Fig5}b).
Their distribution around mean values $F_{max}$ and $F_{min}$ is
nearly gaussian.\\
The dynamical evolution $F(t)$ of the pushing force during the
slip phase can be translated into a function of the granular
material instantaneous velocity. Motion of the granular material
at velocity $V$ induces a variation
$\text{d}F=k(V_{0}-V)\text{d}t$ of resistance force during time
$\text{d}t$, where $k$ is the force probe stiffness and $V_{0}$
the driving velocity; we thus obtain
$V(t)=V_{0}-\text{d}F/k\text{d}t$. We see (Fig. \ref{Fig12}) that
the acceleration and deceleration phases are nearly symmetric.
This phenomenology is currently observed in solid friction
experiments. This is in contrast with previous plane shearing
experiments of granular materials \cite{Gollub97} in which the
deceleration phase would occur in two steps, first fast and then
slow. However, a major difference is that in the latter case, the
displacements during slippage are millimetric (a grain size)
whereas in our experiment, they are micrometric ($30\ \mu$m on
Fig. \ref{Fig12}). As a consequence, the evolution of forces
observed during a slip seems to reflect the slippage of grain
contacts at the walls. The maximum velocities obtained during the
slip phase are of a few hundred $\mu\mbox{m.s}^{-1}$. We did not
study systematically the variation of instantaneous velocity with
bead height nor with driving velocity.

In the following, we report on the $F_{max}$ and $F_{min}$ properties as a
function of the packing height, the driving velocity and the relative humidity.

\subsubsection{Mechanical properties}

\begin{figure}
\includegraphics[width=5cm,angle=-90]{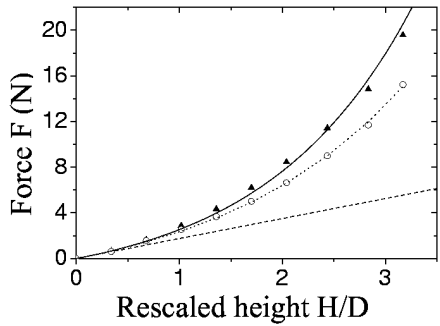}
\caption{Mean maximum and minimum resistance forces $\F$ (filled
triangles) and $\f$ (open circles) in the stick-slip regime as a
function of the height $H$ of the packing scaled by the column
diameter $D$, for steel beads in the duralumin column, for
stick-time $\tstick=40$ s. The full and dotted lines are fit by
eq. (\ref{3}). The dashed line is the hydrostatic curve.}
\label{Fig13}
\end{figure}

In the stick-slip domain, the mean maximum and minimum resistance
forces $F_{max}$ and $F_{min}$ are measured as functions of the
packing height (see Fig. \ref{Fig13}), but now we choose to
perform our experiments not at a constant velocity $V$, but at
constant stick time $t_{stick}$ and we explain why in the
following.

Experimentally we find that the resistance forces $F_{max}$ and
$F_{min}$ increase exponentially with the packing's height $H$.
The data can actually be fitted by a formula of type eq.
(\ref{1}), but the parameter $p_{+1}$ extracted may be different
and should not depend on $\mu_{d}(V_{0})$ as the granular material
is not sliding during stick, and does not slide at driving
velocity $V_{0}$ during slip.

The model can actually be simply modified for this situation. In
the stick-slip regime, when the resistance force $F(t)$ is equal
to $F_{max}$, the material just starts slipping, which means that
static friction is fully mobilized at this instant for each
contact. So we can write $\sigma _{rz}=\mu_{s}\sigma_{rr}$
everywhere at the walls, where $\mu_{s}$ is static friction
between beads and the cylinder's walls. Static friction
coefficients are known to evolve in time \cite{Baumberger97}, we
then have to include aging effects. As the granular material was
at rest in the column since the previous slip, aging time in this
situation is the time of stick:
\begin{equation}
t_{stick}=\frac{F_{max}-F_{min}}{kV_{0}}
\end{equation}
So the final formula for $F_{max}$ in the Janssen framework is:
\begin{equation} F^{\epsilon}=\varrho g\lambda\pi
R^{2}\times\epsilon(\exp(\epsilon \frac{H}{\lambda})-1) \label{3}
\end{equation}
where the screening length is now
$\lambda=R/2K\mu_{s}(t_{stick})$. Here again, $\epsilon=+1$ for a
pushing experiment (i.e. $\F=F^{+1}$), $\epsilon=-1$ for a Janssen
experiment. The choice of imposing a constant stick-time
$t_{stick}$ instead of constant driving velocity $V_{0}$ for the
experiments in the stick-slip regime is now justified by the
dependence of $\mu_{s}$ on $t_{stick}$, whereas $\mu_{d}$ depends
on $V_{0}$.

In the steady state regime, the experimental data obtained for a
given stick-time $t_{stick}$ can be fitted by relation (\ref{3})
by adjusting only one parameter i.e.
$p_{+1}=K\times\mu_{s}(t_{stick})$ where $\mu_{s} (t_{stick})$ is
the static coefficient of friction for an aging time $t_{stick}$.
For the data of Fig. \ref{Fig13} obtained for a relative humidity
$\chi=40\%$ with a mean stick time of $40\ $s, we find
$p_{+1}=0.184\pm0.002$ from $F_{max}$ ($F^{+1}$ in eq. (\ref{3}).
Note that, though friction is not fully mobilized when $F=\f$, the
fit of $\f$ with eq. (\ref{3}) is good (Fig. \ref{Fig13}); we
obtain a parameter $p_{+1}=0.148\pm0.002$ which should not be
related to $\mu_{s} (t_{stick})$ in that case.

\begin{figure}
\includegraphics[width=5cm,angle=-90]{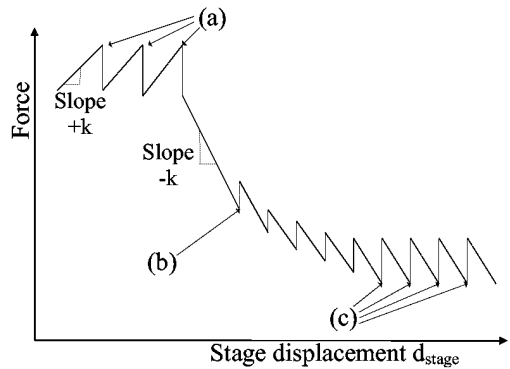}
\caption{Sketch of the method for measuring forces at the onset of
slipping in the pushing and pulling situations. (a) refers to the
maximum force before a slip in the stationary stick-slip regime in
the pushing experiment; (b) refers to the minimum force before the
first slip in the pulling experiment after the material have been
pushed over 5 mm; (c) refers to the minimum force before a slip in
the stationary stick-slip regime in the pulling experiment.}
\label{Fig14}
\end{figure}

As a check of consistency, we perform the following dynamical
experiment in the stick-slip regime. First, the granular column is
pushed upwards in order to mobilize the friction forces downwards
and to reach the steady state compacity; $F_{max}$ (referred to as
(a) on Fig. \ref{Fig14}) is then measured for constant stick time
$t_{stick}$. Starting from this situation, the friction forces are
reversed at the walls by moving the piston downwards. We then
measure the force $F_{min}$ of the first slip (referred to as (b)
on Fig. \ref{Fig14}) after the same time $t_{stick}$. Note that
the roles of $F_{min}$ and $F_{max}$ are inverted in the pulling
experiment: $F_{min}$ is the minimum force before a slip, and
$F_{max}$ is the maximum force at the end of a slip. We also
measure mean minimum force $F_{min}$ (referred to as (c) on Fig.
\ref{Fig14}) when a steady stick-slip regime is reached in the
pulling experiment. Note that this stationary regime is
characterized by the same compacity
$\overset{\_}{\nu}\approx62.5\%$ as in the pushing situation.

\begin{figure}
\includegraphics[width=7cm]{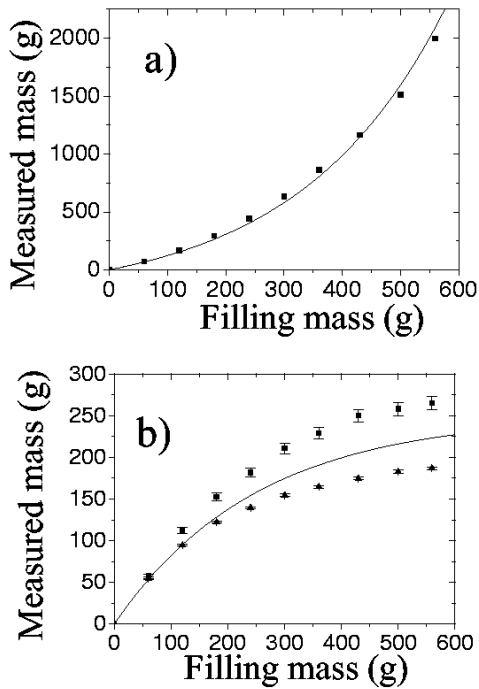}
\caption{a: Mean maximum resistance force measured in the
stick-slip regime as a function of the height $H$ of the packing
scaled by the column diameter $D$, for steel beads in the
duralumin column, for stick-time $\tstick=40$ s; the line is a fit
with eq. (\ref{3}). b: Measured mass vs. filling mass for a
classical Janssen experiment obtained in two different ways: we
plot the mass of the first slipping event (squares), and the mean
minimum mass in the stationary stick-slip for the pulling
situation (triangles); the line is the Janssen curve predicted
with the parameter extracted from the pushing experiment of Fig.
\ref{Fig15}a.} \label{Fig15}
\end{figure}

Following relation (\ref{3}), this procedure should imply a change
of $\epsilon$ from $1$ to $-1$, and consequently, the force at the
onset of slipping on the piston should decrease from $F^{+1}$ to
$F^{-1}$. As we observed no noticeable variation of compacity
between the pushing and the pulling, we expect Janssen's parameter
$K$ value to be unchanged. In Fig. \ref{Fig15} the pulling force
$F^{-1}$ is measured for different packing heights $H$, for the
first slipping event and in the steady stick-slip regime. The fit
of experimental results with eq. (\ref{3}) gives $p_{-1}(40\ \mbox
{s})=0.230\pm0.004$ for stationary stick-slip which is $20\%$
larger than $p_{+1}(40\ \mbox{s})$. It cannot be due to a slight
change in average compacity $\overset{\_}{\nu}$ as from relation
$\Delta K/K\approx 5\Delta\overset{\_}{\nu}/\overset{\_}{\nu}$, we
would expect a $4\%$ variation in compacity, which would have been
observed; let us recall that we actually measured a compacity
variation between the pushing and the pulling experiment $\Delta\overset{\_}{\nu}/\overset{\_}{\nu}=0\pm1\%$.\\
As in the steady-sliding experiment, according to Janssen's
picture, this would then imply that vertical stress redirection is
more efficient in the pulling situation; furthermore, this effect
is enhanced in the stick-slip regime. For the first slip event in
the pulling experiment, we find $p_{-1}(40\
\mbox{s})=0.16\pm0.015$ which is now lower than $p_{+1} (40\
\mbox{s})$. We believe this is a clear evidence of a granular
structuration effects. We interpret the differences in $p_{-1}$
values by saying that the packing has first been structured to
resist the pushing; when we start pulling, structure is not
efficient to resist the pulling, and it results in a lower
Janssen's constant (therefore lower $p_{-1}$) for the first slip
event; when we continue pulling, the packing gets structured to
resist pulling and leans more efficiently on the walls, i.e.
Janssen's constant (and $p_{-1}$) increases. Moreover,
steady-state structuration is more efficient for the pulling than
for the pushing, which means that both structurations may be
different and could reflect the symmetry breaking due to gravity.

The results of the simulations of an elastic medium in a column
presented in the precedent section for a steady-sliding also carry
on to the stick-slip situation. If we impose numerically
everywhere at the walls, the relations
$\sigma_{rz}=\boldsymbol{\mu}_{\mathbf{s}}\sigma_{rr}$ or $\sigma
_{rz}=\boldsymbol{\mu}_{\mathbf{d}}\sigma_{rr}$, it would merely
correspond to a change in the name of the friction coefficient.
The equilibrium equations are the same for the static and steady
dynamic cases. The important common point is that friction forces
are fully mobilized everywhere at the walls. Therefore, as for the
steady-sliding, the data can be fitted by the elastic predictions,
and the $3\%$ difference in Janssen's parameter between the
pulling and the pushing of an elastic medium of constant Poisson
ratio is again qualitatively (though not quantitatively) in
agreement with the experimental results.

In the following, as we study the steady-state regime, we will try
to distinguish between wall-bead interactions (i.e. $\mu_{s}$) and
the bulk properties (i.e. $K$) as they influence the rheological
properties when velocity and relative humidity are changed.

\subsubsection{Rheological properties}

\begin{figure}
\includegraphics[width=7cm]{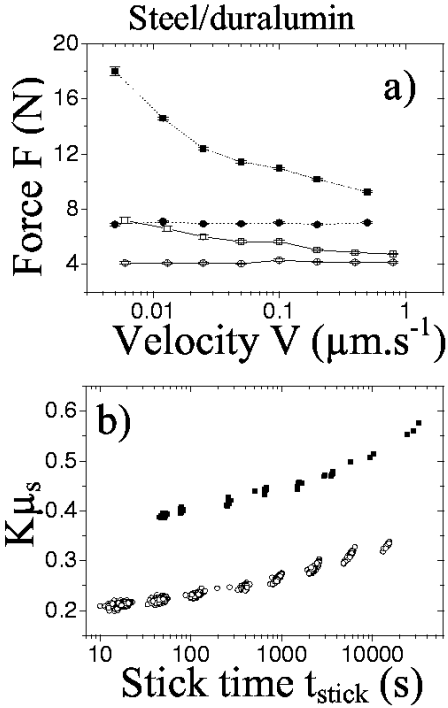}
\caption{a: Minimum and maximum resistance forces $\f$ (circles)
and $\F$ (squares) vs. driving velocity in the stick-slip regime
for a height $H=1.4D$ (250 g) of steel beads in the duralumin
cylinder, for relative humidities $\chi=90\%$ (filled symbols) and
$\chi=45\%$ (open symbols). b: $K\mu_s$ as a function of stick
time $\tstick$ for a height $H=1.4D$ (250 g) of steel beads in the
duralumin cylinder, for relative humidities $\chi=90\%$ (filled
squares) and $\chi=45\%$ (open circles).} \label{Fig16}
\end{figure}

\begin{figure}
\includegraphics[width=7cm]{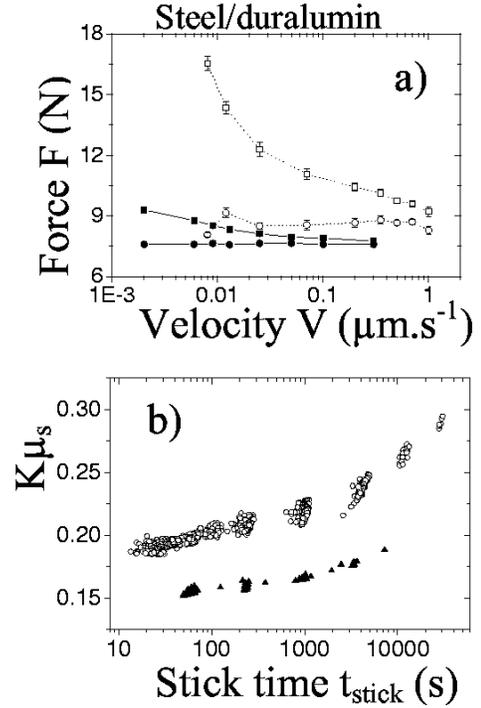}
\caption{a: Minimum and maximum resistance forces $\f$ (circles)
and $\F$ (squares) vs. driving velocity in the stick-slip regime
for a height $H=2.15D$ (380 g) of steel beads in the duralumin
cylinder, for relative humidities $\chi=45\%$ (open symbols) and
$\chi<3\%$ (filled symbols). b: $K\mu_s$ as a function of stick
time $\tstick$ for a height $H=2.15D$ (380 g) of steel beads in
the duralumin cylinder, for relative humidities $\chi=45\%$ (open
circles) and $\chi<3\%$ (filled triangles).} \label{Fig17}
\end{figure}

\begin{figure}
\includegraphics[width=7cm]{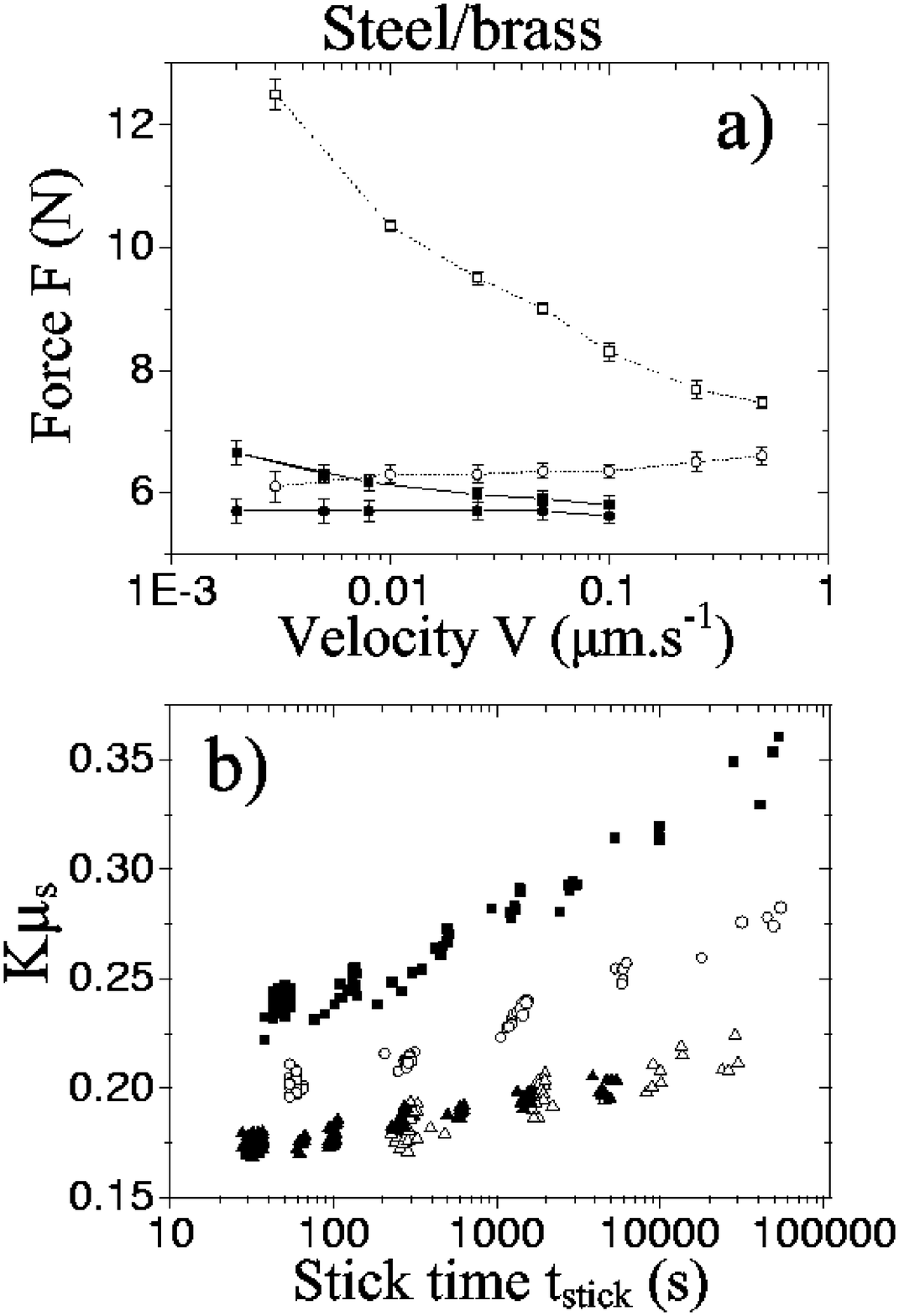}
\caption{a: Minimum and maximum resistance forces $\f$ (circles)
and $\F$ (squares) vs. driving velocity in the stick-slip regime
for a height $H=1.7D$ (300 g) of steel beads in the brass
cylinder, for relative humidities $\chi=90\%$ (open symbols) and
$\chi<3\%$ (filled symbols). b: $K\mu_s$ as a function of stick
time $\tstick$ for a height $H=1.7D$ (300 g) of steel beads in the
brass cylinder, for relative humidities $\chi=90\%$ (filled
squares) and $\chi<3\%$ (filled triangles), and for a height
$H=2.15D$ (380 g) for relative humidities $\chi=64\%$ (open
circles) and $\chi<3\%$ (open triangles).} \label{Fig18}
\end{figure}

The mean maximum force $F_{max}$ increases strongly when the
driving velocity $V_{0}$ is decreased and the relative humidity
$\chi$ is increased (Fig. \ref{Fig16}a, \ref{Fig17}a,
\ref{Fig18}a). Furthermore, $F_{max}$ apparent divergence when
$V_{0}$ decreases is enhanced when $\chi$ increases. The mean
minimum force $F_{min}$ is almost independent of the driving
velocity, and increases when the relative humidity is increased
(Fig. \ref{Fig16}a, \ref{Fig17}a, \ref{Fig18}a). For each
stick-slip event, an aging time (the stick time) is measured:
$t_{stick} =(F_{max}\!-\!F_{min})/(kV)$. This time is associated
to a parameter $K\mu _{s}(t_{stick})$ extracted from the fit of
the maximum force $F_{max}$ of this stick-slip event with equation
(\ref{3}). In the following, the results of this analysis for
duralumin and brass columns are compared and discussed.

\subsubsection*{\textit{Steel/duralumin}}

For steel beads in the duralumin column, the parameter $K\mu_{s}$
(Fig. \ref{Fig16}b, \ref{Fig17}b) increases roughly
logarithmically with aging time, and increases strongly with
relative humidity $\chi$. Its value increases by 50\% when $\chi$
increases from $3\%$ to $45\%$, and doubles when $\chi$ increases
from $45\%$ to $90\%$. The parameter $K\mu_{s}$ does not depend
only globally on $\chi$, but its logarithmic aging rate $Kb$ with
stick time $t_{stick}$ depends strongly on $\chi$. The parameters
$Kb$ extracted from a logarithmic fit of $K\mu_{s}(t)$ with
$\mu_{s}(t)=\mu_{0}+b\ln(t/\tau_{0})$ are given in table
\ref{tablevieillissementdural} for different relative humidities.
\newline Note that magnitude of $Kb$, as $K$ is of
the order 1, is consistent with many previous reports on solid
friction \cite{Berthoud99,Crassous99}. An important feature is
that the logarithmic aging rate appears to be 4 times higher in
humid atmosphere than in dry atmosphere.

\begin{table}
\begin{center}
\begin{tabular}
[c]{|c|c|}\hline relative humidity $\chi$ & $Kb$\\\hline
$<3\%$ & $0.006\pm0.001$\\
$45\%$ & $0.013\pm0.003$\\
$90\%$ & $0.023\pm0.001$\\\hline
\end{tabular}
\end{center}
\caption{Parameter $Kb$ extracted from a fit of $K\mu_{s}(t)$ with
$\mu _{s}(t)=\mu_{0}+b\ln(t/\tau_{0})$ for different relative
humidities $\chi$, for a height $H=1.4D$ (250 g) and a height
$H=2.15D$ (380 g) of steel beads in the duralumin column.}
\label{tablevieillissementdural}
\end{table}

The logarithmic evolution is not perfect: for aging time larger
than $3000\ $s, we observe that $K\mu_{s}$ increases more sharply
with time; for $H=2.15D$ of steel beads and $\chi=45\%$, $Kb$
varies from $0.009$ for times $t_{stick}<3000$ s to about $0.025$
in the last time decade (Fig. \ref{Fig17}b). This phenomenon was
observed for all relative humidities $\chi$. It is similar to what
was reported previously by Losert et al. \cite{Gollub97} for plane
shearing of glass beads.\\ It is tempting to link this enhanced
aging behavior to the granular material slow restructuring effects
since one may think of an aging of the granular structure leading
to a better stress redirection at the walls.

However, at this point, it is not possible to differentiate
between wall and structure properties. We tried to measure
independently $\mu_{s}$ aging properties in a solid on solid
experiment, but we did not obtain satisfactory statistics to
conclude. Therefore, we need to compare the results obtained in
the duralumin column to the one obtained in the brass column.

\subsubsection*{\textit{Steel/brass}}

For steel beads in the brass column, the parameter $K\mu_{s}$
extracted from the experimental measurements (Fig. \ref{Fig18}b)
also increases logarithmically with aging time. However, the
increase with relative humidity $\chi$ is much less important than
in the duralumin column since $K\mu_{s}$ increase is only $50\%$
when increasing $\chi$ from dry atmosphere to $90\%$ humidity,
whereas the increase is more than $150\%$ in the duralumin column
for the same variation of $\chi$. Moreover, the $K\mu_{s}$
increase with stick time seems now perfectly logarithmic over 4
decades of time variation.

As in the duralumin column, $K\mu_{s}$ increase rate with stick
time $t_{stick}$ depends on $\chi$. The parameters $Kb$ extracted
from a logarithmic fit of $K\mu_{s}(t)$ with
$\mu_{s}(t)=\mu_{0}+b\ln(t/\tau_{0})$ are given in table
\ref{tablevieillissementlaiton}. Logarithmic aging rate is now 2.5
times greater in humid atmosphere than in dry atmosphere (it was 4
times higher in the duralumin column).

\begin{table}
\begin{center}
\begin{tabular}
[c]{|c|c|}\hline relative humidity $\chi$ &  $Kb$\\\hline
$<3\%$ & $0.006\pm0.001$\\
$64\%$ & $0.011\pm0.0005$\\
$90\%$ & $0.014\pm0.0005$\\\hline
\end{tabular}
\end{center}
\caption{Parameter $Kb$ extracted from a fit of $K\mu_{s}(t)$ with
$\mu _{s}(t)=\mu_{0}+b\ln(t/\tau_{0})$ for different relative
humidities $\chi$, for a height $H=1.7D$ (300 g) and a height
$H=2.15D$ (380 g) of steel beads in the brass column.}
\label{tablevieillissementlaiton}
\end{table}

\subsubsection*{\textit{Comments}}

The only differences between both systems (duralumin and brass
column) being the contacts at the walls, as for the steady-sliding
regime, the differences observed in the rheological behavior
suggest that the aging of $K\mu_{s}$ is mainly an aging of
friction coefficient $\mu_{s}$ at the walls and that humidity
affects principally the contacts at the walls.

Relative humidity $\chi$ seems to have two effects in the
duralumin column: first, the aging rate is higher when $\chi$ is
higher, second, the friction level for short times is higher when
$\chi$ is higher.\newline In the brass column there seems to be an
effect of $\chi$ only on the aging rate: the $K\mu_{s}(t_{stick})$
curves for different $\chi$ cross for $t=0.1$ ms. The logarithmic
aging rate coefficient gets comparable values in both columns, but
it is higher in the duralumin column ($Kb=0.023$) in humid
atmosphere than in the brass column ($Kb=0.014$).

Our observations are consistent with recent aging experiments in
granular media \cite{Boquet98,Restagno02} and solid on solid aging
experiments \cite{Crassous99}, who found logarithmic aging of
static friction enhanced by an increase in relative humidity
$\chi$. These results have been interpreted by the dynamics of
capillary condensation at the contacts \cite{Boquet98,Restagno00},
which is a thermally activated process: as time goes on, there are
more and more capillary bridges at the contacts, which are
responsible for an adhesion force increasing with aging time;
moreover, condensation goes faster when $\chi$ is higher. In these
models, there is no aging in dry atmosphere, as no condensation
can occur; this prediction agrees with most observations
\cite{Crassous99}. As in our experiment the logarithmic aging rate
coefficient is not zero in dry atmosphere (even though the air
flux we impose is probably not perfectly dry), we can conclude
that capillary condensation is not the only source of aging.
Another source of aging can be creep of contacts
\cite{Berthoud99}.

We observed that aging is perfectly logarithmic in the brass
column. Therefore, we are led to analyze again the aging in the
duralumin column since the interpretation based on aging of
internal friction we proposed for the duralumin column, should
also apply to the brass column, which is in contradiction with the
observed behavior.

If we suppose that $K$ properties depend only on the granular
material properties, the acceleration of logarithmic aging in the
duralumin column may be not linked to aging of bead/bead contacts
nor granular slow restructurations. We may then start from the
following experimental observations: in solid on solid friction
experiments, static friction coefficients $\mu_{s}$ are found to
depend on the applied shear exerted for a given waiting time
\cite{Berthoud99,OvarlezGauthier}. For a higher shear during a
given waiting time, coefficient $\mu_{s}$ is higher, and the
logarithmic aging rate is higher too \cite{Berthoud99}. In our
experiments, during a ``stick'' event, the pushing force increases
linearly with time, i.e. shear at the walls (and contact pressure)
increases. As a consequence, aging occurs with a non-constant
applied shear and the mean shear is more important when the aging
time is higher. Consequently for high waiting time or stick time,
i.e. for high mean shear at the walls, the aging of the static
friction coefficient may be accelerated.\newline This
phenomenology may be a priori different for different surfaces,
which would explain why aging seems perfectly logarithmic in the
brass column, and not in the duralumin column.

\subsubsection{Analysis of $F_{max}$ evolution}

Now, the high enhancement of the blocking resistance force $F_{max}$ with
decreasing driving velocity can be simply understood within the Janssen
framework . We start from the experimental observation that the minimum force
$F_{min}$ after a slip does not depend on velocity (at constant relative
humidity $\chi$): $F_{min}=F_{0}$. Therefore, the aging time $t_{stick}$
before a slip reads:
\begin{equation}
t_{stick}=(F_{max}\!-\!F_{0})/(kV)
\end{equation}
We suppose, as well verified experimentally, that the coefficient
of friction $\mu_{s}$ at the walls follows a logarithmic aging
law:
\begin{equation}
\mu_{s}(t)=\mu_{0}+b\ln(t/\tau)
\end{equation}
This value of $\mu_{s}(t)$ for aging time $t_{stick}$ can be
injected in formula (\ref{3}). We then obtain:
\begin{equation}
F_{max}\propto V^{-1/\gamma}
\end{equation}
with $\gamma=R/(2KbH)-1$.\newline The apparent divergence of
$F_{max}$ with decreasing velocity, as observed experimentally, is
thus more important when the height of beads and the logarithmic
aging rate $b$ of friction at the walls (i.e. relative humidity as
$b$ increases with $\chi$) are higher. The observed strong
dependence of $F_{max}$ with $V$ then corresponds to an
exponential amplification of friction aging at the walls. Elements
of interpretation along these lines were already given in ref
\cite{Ovarlez01}.

\subsection{Influence of relative humidity: an open problem}

Nevertheless from the whole series of experiments we performed on this system
we are led to conclude that the dependence of the phenomenology with the
relative humidity $\chi$ remains sometimes unclear.

\begin{figure}
\includegraphics[width=7cm]{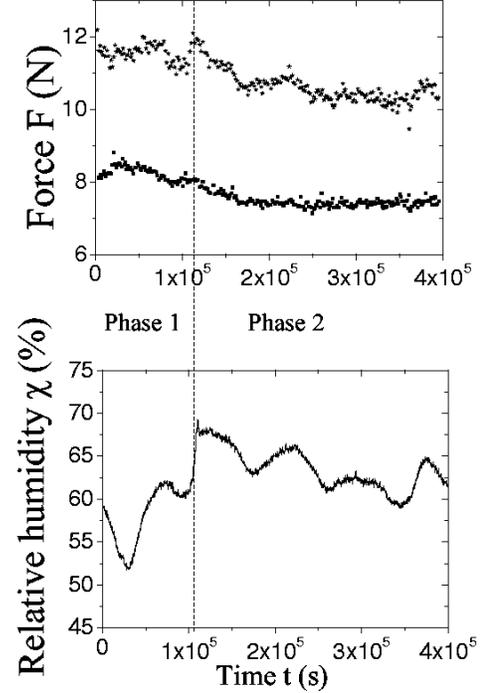}
\caption{a: Maximum and minimum resistance forces to pushing $\F$
(stars) and $\f$ (squares) of a height $H= 2.15D$ (380 g) of steel
beads in the duralumin cylinder vs. time in the stick-slip regime
($V=50\ \mbox{nm.s}^{-1}$). b: Evolution of relative humidity
$\chi$ during the experiment of Fig. a. The dotted line separates
two phases: 1) forces and $\chi$ evolutions are not correlated; 2)
forces and $\chi$ evolutions are correlated.} \label{Fig19}
\end{figure}

\begin{figure}
\includegraphics[width=5cm,angle=-90]{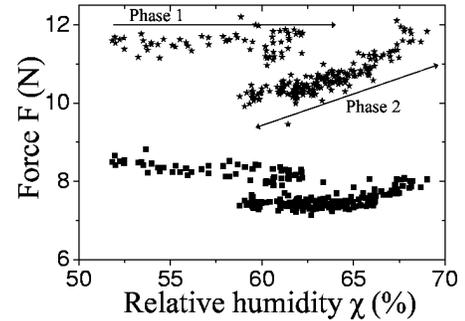}
\caption{Maximum and minimum forces $\F$ (stars) and $\f$
(squares) vs. relative humidity $\chi$ for the experiment of Fig.
\ref{Fig19}; phases 1) and 2) of Fig. \ref{Fig19} are indicated on
the graph.} \label{Fig20}
\end{figure}

As an illustration of this statement, we report the following
result (Fig. \ref{Fig19}, \ref{Fig20}). In an experiment performed
in the stick-slip regime and in the ambient atmosphere, for a
driving velocity $V=50\ \mbox{nm.s}^{-1} $, the forces $F_{min}$
and $F_{max}$ were found to be independent of $\chi$, which was
varying between $52\%$ and $62\%$, during the first 30 hours of
the experiment. We then observed that the $F_{min}$ and $F_{max}$
variations became suddenly correlated to the $\chi$ variations and
would stay correlated for the next 80 hours. For the first 30
hours, $\chi$ increased from $52\%$ to $62\%$, while $F_{min}$ and
$F_{max}$ were nearly constant: $F_{max}=11.6$ N, $F_{min}=8.2$ N.
Then $\chi$ varied from $62\%$ to $68\%$ in one hour: $F_{min}$
and $F_{max}$ started to be correlated to $\chi$ from this ``
triggering event'' and for the next 80 hours as if the system had
suddenly reached some ``reversible branch''.

Such a phenomenology was never obtained when humidities $\chi=90\%$ and
$\chi<3\%$ were imposed. When we imposed dry atmosphere, the force would reach
its stationary value in a few minutes, and we never noticed any variation in
this value for 5 days; when we imposed humid atmosphere, the force would reach
its stationary value in a maximum of 3 hours and we noticed no variation in
this value for 5 days.

This may however incite to prudence for results obtained in
ambient atmosphere, for experiments performed on the granular
column as much as for solid on solid friction experiments. As an
example, we see on Fig. \ref{Fig20} that for $\chi=60\%$,
$F_{max}=11.5\pm0.2$ N when forces and $\chi$ are not correlated,
whereas $F_{max}=10.3\pm0.2$ N when they are correlated.

Furthermore, this opens the question of the metastability of capillary
condensation. If variation of $F_{max}$ with $\chi$ is interpreted as
variation of $\mu_{s}$ due to thermally activated capillary condensation at
the contacts \cite{Boquet98,Restagno00}, it seems possible that capillary
condensation did not occur for hours. If we notice that $F_{max}$ starts to be
correlated with $\chi$ when $F_{max}$ constant value is equal to $F_{max}$ on
the $\chi$ dependent part of the curve (i.e. $F_{max}=11.8$ N and $\chi
=68\%$), another possibility is that there were as much capillary bridges
condensed at the contacts as for $\chi=68\%$ in the first part of the
experiment, i.e. that evaporation does not always happen.

\subsection{Dieterich-Ruina analysis}

A standard phenomenological model accounting for static and dynamic properties
of solid friction is the so-called Dieterich-Ruina model
\cite{Dieterich79,Ruina}. In this picture, the coefficient of friction is
\begin{equation}
\mu=\mu_{0}+a\ln(\frac{V}{V_{0}})+b\ln{\frac{V_{0}\theta}{D}}
\end{equation}
where $\theta$ obeys \cite{othermodels}
\begin{equation}
\frac{\mbox{d}\theta}{\mbox{d}t}=1-\frac{\theta V}{D}
\end{equation}
$\mu_{0}$, $a$, $b$, and $D$ are constants dependent of the
materials in contact. Parameter $D$ is usually interpreted as a
characteristic length for renewal of contacts and is of order of
the creep length before sliding \cite{Heslot94}. In a stationary
steady-sliding regime $\theta=D/V$ is interpreted as a
characteristic time for renewal of contacts.

This model accounts for the logarithmic aging of static friction, and the
logarithmic velocity strengthening or weakening in the steady-sliding regime
with:
\begin{align}
\mbox{d}\mu_{s}/\mbox{d}\ln(t)  &  =b\\
\mbox{d}\mu_{d}/\mbox{d}\ln(V)  &  =a-b
\end{align}
It also accounts for a third usual phenomenon observed in the transitory
regime when the velocity is changed suddenly. In this case there is an abrupt
change in friction coefficient followed by exponential relaxation to a new
stationary value. If we change velocity $V_{1}$ to velocity $V_{2}$, the
instantaneous change in friction coefficient from stationary value $\mu
_{d}(V_{1})=\mu_{0}+(a-b)\ln(V_{1}/V_{0})$ is $+a\ln(V_{2}/V_{1})$; friction
change from this value to new stationary value $\mu_{d}(V_{2})$ is then
$-b\ln(V_{2}/V_{1})$ during typical time $\tau=D/V_{2}$, i.e. we get $\mu
_{d}(V_{2})=\mu_{0}+(a-b)\ln(V_{2}/V_{0})$.

Now we test the Dieterich-Ruina model with the steel/duralumin and steel/brass
friction coefficient extracted from our data. As we observed logarithmic aging
and rough logarithmic velocity strengthening, we get parameters $a$ and $b$.
We are therefore able to test this model predictions with the observation of
response to an abrupt variation of velocity.

Experimentally, we never observed any transitory state: the force
in the steady sliding regime was always changed from stationary
value $\bar{F} (V_{1})$ to stationary value $\bar{F}(V_{2})$ when
changing velocity from $V_{1}$ to $V_{2}$. From the measured value
of $Kb=0.010$ for steel/duralumin friction and $Kb=0.011$ for
steel/brass in ambient atmosphere, we would however expect the
force at the beginning of the transitory regime to be $0.5\ $N
larger (steel beads height $H=2.15D$) than the stationary force at
the new velocity. As the natural fluctuations of force are around
$0.1\ $N, we would therefore expect to observe this transitory
regime in $\bar{F}$ due to change in friction at the walls. Now we
give an interpretation for why, within the Dietrich-Ruina picture,
this transitory regime was not observed, which yields an upper
bound for the material contact renewal length $D_0$.

When velocity is changed from $10\ \mu\mbox{m.s}^{-1}$ to $100\
\mu\mbox {m.s}^{-1}$ for $H=2.15D$ (380 g) of steel beads in a
duralumin cylinder, the change in $\bar{F}$ stationary value is of
1 N. But, due to the finite stiffness of our set-up, this change
cannot be instantaneous: an increase $\Delta F$ of force at
constant velocity takes minimal time $t=\Delta F/(kV)$. So an
increase of 1 N of force at $100\ \mu\mbox{m.s}^{-1}$ takes time
$t=0.25\ $s. This time must be compared to time length $\tau$ of
transitory regime: $\tau=D_{0}/V$. Therefore, in the framework of
the Dieterich-Ruina model, we need $\tau$ to be less than the time
of force variation, which means $D_{0}<25\ \mu$m .

Let us now consider a sudden decrease in velocity from $100\
\mu\mbox {m.s}^{-1}$ to $1\ \mu\mbox{m.s}^{-1}$. Now the decrease
of force is governed by inertial time: $t_{in}=\pi\sqrt{m/k}$
which can be evaluated around $t_{in}=200$ ms. This leads to
$D_{0}<400$ nm (so that $\tau<t_{in}$) consistently with the
Dieterich-Ruina model if no transitory state is observed.

We have no precise measure for $D_{0}$ but this upper limit on the
value is coherent with the usual interpretation in terms of length
for contact renewal, as the duralumin column mean roughness is 400
nm. Thus, we can neither validate nor rule out the Dieterich-Ruina
model for our system.

\subsection{Transition mechanism}

\begin{figure}
\includegraphics[width=5cm,angle=-90]{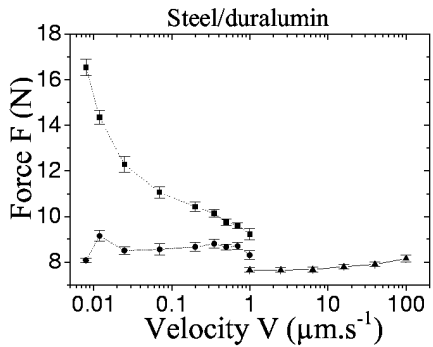}
\caption{Mean maximum and minimum forces $\F$ (squares, dotted
line) and $\f$ (circles, dashed line) in the stick-slip regime,
and mean resistance force $\fs$ (triangles, full line) in the
steady-sliding regime for a height $H= 2.15D$ (380 g) of steel
beads in the duralumin cylinder, for relative humidity
$\chi=40$\%.} \label{Fig21}
\end{figure}

For experiments performed at given height of beads and relative
humidity $\chi$, we observe that the transition from stick-slip to
steady-sliding occurs for a finite stick-slip amplitude (Fig.
\ref{Fig21}). This suggests a subcritical transition but such a
transition would be hysteretic. In order to verify this feature,
we make the following experiment: we drive the system at constant
velocity in the steady-sliding regime near the transition until it
reaches a stationary state, we then decrease continuously the
velocity, without ever stopping the movement, to a velocity for
which stick-slip used to occur. Note that the results reported in
the previous sections and on Fig. \ref{Fig21} were obtained for
independent experiments on systems driven at a unique velocity. We
observe that if we impose $V=700\ \mbox{nm.s}^{-1}$ directly, we
obtain a stick-slip motion, whereas when we decrease velocity from
$1.4\ \mu\mbox{m.s}^{-1}$ (in the steady-sliding regime) to
$V=700\ \mbox{nm.s}^{-1}$ the motion is steady-sliding. This
proves the hysteretic character of the transition, which was
actually hard to observe as the system seems very sensitive to
noise.

This transition is similar to the one observed by Heslot et al.
\cite{Heslot94} in solid on solid experiments; it correspond to
what they call the ''inertial regime''. It is a characteristic of
systems for which the dynamical coefficient of friction increases
with velocity. This transition scenario was explained by Brockley
et al. \cite{Brockley}. However they find that the force $\bar{F}$
in the steady-sliding regime at the transition is of order of
$F_{min}$. We actually find (see Fig. \ref{Fig21}) that $\bar{F}$
is less than $F_{min}$ for steel beads in the duralumin cylinder,
a feature we do not understand, whereas $\bar{F}\approx F_{min}$
for in the brass cylinder.

Again, the transition we observed is consistent with a picture of solid
friction sliding instability at the wall. This is also consistent with the
model we have developed in the previous section.

\section{Polydisperse glass beads}

After the study of the simplest case, with monodisperse low
friction beads, we now want to study the effect of disorder
(friction, polydispersity) on rheology. Therefore, we choose to
study a miscellany of glass beads using a three diameters mixture
(1.5 mm, 2 mm and 3 mm) with equal volume of each kind in an
abraded PMMA cylinder. In the following, we present the features
obtained in the stick-slip regime for this system.

\subsection{Stick-slip characterization}

\begin{figure}
\includegraphics[width=7cm]{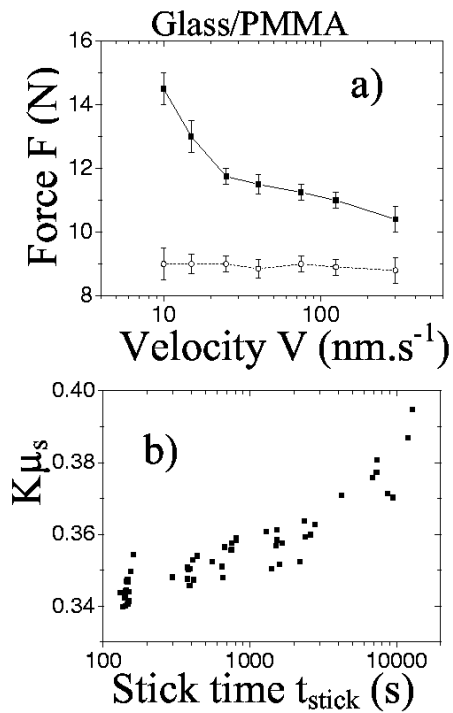}
\caption{a: Minimum and maximum resistance forces $\f$ (open
circles) and $\F$ (full squares) vs. driving velocity in the
stick-slip regime for a height $H=2.8D$ (132 g) of a miscellany of
glass beads in the PMMA cylinder at relative humidity $\chi=35$\%.
b: $K\mu_s$ as a function of stick time $\tstick$ extracted from
$\F$ with equation (\ref{3}).} \label{Fig22}
\end{figure}

\begin{figure}
\includegraphics[width=7cm]{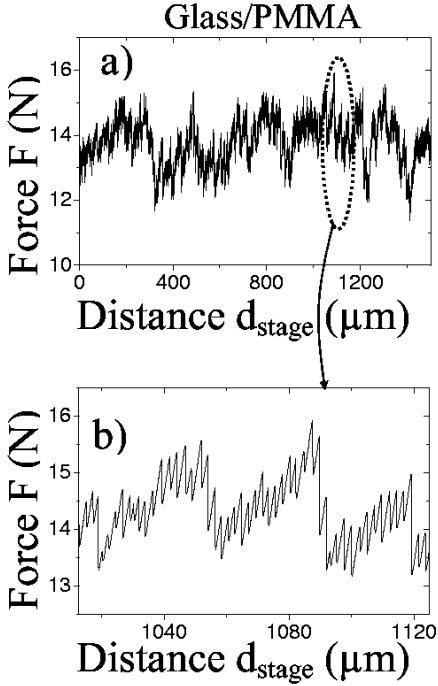}
\caption{a: Resistance force to pushing of a height $H=2.7D$ (126
g) of a miscellany of glass beads in the PMMA cylinder vs.
displacement $d_{stage}$ of the translation stage in the
stick-slip regime ($V=1500\ \mbox{nm.s}^{-1}$). b: zoom of Fig.
\ref{Fig23}a.} \label{Fig23}
\end{figure}

For low velocities ($V<100\ \mbox{nm.s}^{-1}$), a stick-slip
motion is obtained with the same properties as observed with
monodisperse steel beads in a duralumin or brass cylinder. It is a
regular stick-slip motion characterized by a maximum force before
slipping that increases strongly when velocity is decreased, and a
minimum force after slipping which does not depend on velocity
(Fig. \ref{Fig22}a). A coefficient $K\mu_{s}(t_{stick})$ can be
extracted from $F_{max}$ with equation (\ref{3}) for low
velocities ($V<100\ \mbox{nm.s}^{-1}$). Again, it shows a
logarithmic aging of friction at the walls (Fig. \ref{Fig22}b).
However, the behavior changes for higher velocities just before
the transition to steady-sliding. The stick-slip motion is then
less and less regular as the velocity is increased. Furthermore we
observe structures in the force signal. Figure \ref{Fig23}
evidences cycles of force increase during which the maximum and
minimum forces increase for several stick-slip events, as the
force amplitude $\Delta F_{stick}$ during a stick is higher than
the force amplitude $\Delta F_{slip}$ during a slip. Then, at the
end of a cycle, there is a big slip event and a new cycle starts.
This phenomenon is similar to the one observed by Albert et al.
\cite{Albert99} for the pushing of a stick in a granular material,
and to the one observed for aluminium beads by Kolb et al.
\cite{LaKolb99} in the same display as ours but in 2D . It was
however much more important in that last study in 2D: the force
$F_{max}$ would then increase from $2N$ after a big slip event to
$20N$ before the next big slip event.

\begin{figure}
\includegraphics[width=7cm]{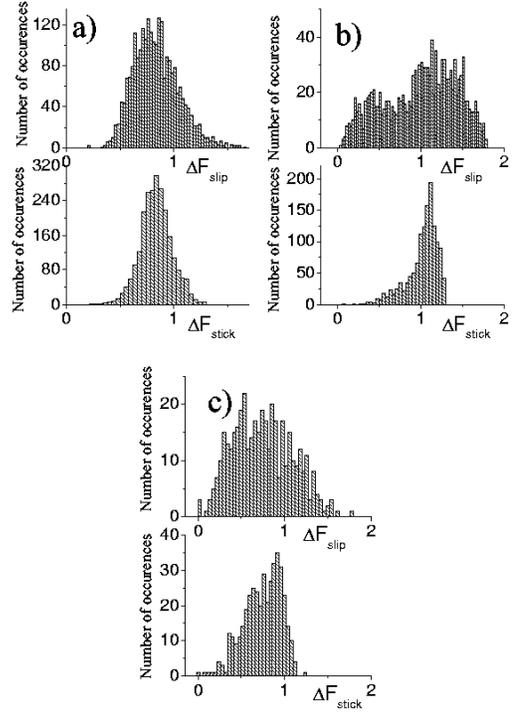}
\caption{Distribution of force variations $\Delta
F_{stick}=\F(i)-\f(i)$ during a stick and $\Delta
F_{slip}=\F(i)-\f(i+1)$ during a slip of a height $H=2.7D$ (126 g)
of a miscellany of glass beads in the PMMA cylinder for three
driving velocities: a) $V=150\ \mbox{nm.s}^{-1}$; b) $V=750\
\mbox{nm.s}^{-1}$; c) $V=1500\ \mbox{nm.s}^{-1}$.} \label{Fig24}
\end{figure}

We analyse this phenomenon by studying the distribution of force
amplitudes defined as $\Delta F_{stick}=F_{max}(i)-F_{min}(i)$ for
the $i^{th}$ stick event, and $\Delta
F_{slip}=F_{max}(i)-F_{min}(i+1)$ for the following slip event
(Fig. \ref{Fig24}). We observe the evolution of these
distributions with increasing velocity. $\Delta F_{stick}$
distribution remains centered around a mean value; however, this
distribution gets larger when the velocity is increased. For lower
velocities ($150\ \mbox{nm.s}^{-1}$ on Fig. \ref{Fig24}a),
stick-slip is regular, and $\Delta F_{slip}$ distribution is
regular around same mean value as $\Delta F_{stick}$ (with
nevertheless a larger distribution). When velocity is increased,
the distribution is still predominant around same mean value as
$\Delta F_{stick}$ but many smaller values start to merge in the
distribution ($V=750\ \mbox{nm.s}^{-1}$ on Fig. \ref{Fig24}b);
these small values become as frequent as the larger ones for
higher velocities ($V=1500\ \mbox {nm.s}^{-1}$ on Fig.
\ref{Fig24}c). The distribution also enlarges to the higher values
which corresponds to occurrence of the big slip events. Note that
in these experiments, the transition to steady-sliding regime
occurs for velocity a velocity around $V=2000\ \mbox{nm.s}^{-1}$.

\subsection{Analysis}

Several explanations can be proposed to account for this complex
dynamics exhibiting a large distribution of energy release. In our
point of view, the principal ingredient for an explanation seems
to be the progressive structuration of the packing between two big
slip events. In the Janssen model picture, the progressive
increase of $F_{max}$ in a cycle corresponds to progressive
increase of Janssen's constant $K$. The $K$ value increases from
an initial value $K_{0}$ at the beginning of a cycle up to a
maximum value $K_{max}$ at the end of the cycle, and is
reinitialised by a big slip event to value $K_{0}$. Let us recall
that $K$ was found to increase with density in another report
\cite{Ovarlez03}. A consequence of this dependence was shown in
section \ref{III.A.1}. Therefore, a way to obtain the
phenomenology is to start from a loose packing at the beginning of
a cycle; if each small slip event densifies the packing (as can be
expected from compaction under vibration), it leads to a small
increase in $K$, so that the minimum force $F_{min}$ at the end of
the slip is higher than the force $F_{min}$ at the end of the
preceding slip; the maximum force $F_{max}$ at the end of the
stick is also higher than the force $F_{max}$ at the end of the
preceding stick. Sometimes, big events occur; they probably
correspond to convective motion at the walls, as observed by Kolb
et al. in 2D \cite{LaKolb99}, which loosens the packing, which
then gets the initial density of the cycle.

For the interpretation we give, the stick events last a constant
time as the aging time of contacts at the walls is still
controlled by the velocity. This explains why the distribution of
$\Delta F_{stick}$ is still centered on mean value contrary to
$\Delta F_{slip}$.

But still a criterion for big slip events has yet to be found; it
may be the level of stresses, or a dilatance effect at high
density causing shear band and convection rolls. The regularity of
stick-slip motion at low velocities can be understood if the
criterion for big slip events is at an upper stress level. A slow
driving velocity leads to long aging of contacts at the walls so
that the maximum force $F_{max}$ is always greater than the
threshold to obtain a big slip. All slip events are then big slip
events in this case.

This transition from a simple steady stick-slip to a complex
dynamical regime with a large distribution of elastic energy
release is an interesting issue and we plan to pursue further the
investigation trying to clarify in particular the role of disorder
and polydispersity in the phenomenology.

\section{Conclusion}

We presented an experimental report on the dynamical behavior of a
granular column pushed vertically. We first investigated the case
of a monodisperse assembly of steel beads and we observed at low driving velocities ($%
1\!<\!V\!<\!100\ \mu m/s$) a steady behavior such that the pushing
forces increase roughly logarithmically with velocity. On the
other hand, at very low driving velocity ($V\!<\!1\ \mu m/s$), we
evidenced a discontinuous and hysteretic transition to a
stick-slip regime characterized by a strong divergence of the
maximal blockage force when the velocity goes to zero. All this
phenomenology is strongly influenced by surrounding humidity:
generally, higher humidity level increases strongly the resistance
to pushing. Finite elements numerical simulations were used to
confront experimental results to a modelling of the granular
packing as an isotropic elastic medium. Then, we showed that a
simple Janssen's model is a fair base for analysis as it provides
the correct physical interpretation for the pushing resistance.
This Janssen's model was used to extract an important mechanical
parameter combining the effects of stress redirection and wall
friction but there is an inherent difficulty to isolate clearly
the various contributions either coming from bulk reorganization
or from the surface friction properties.

Using different column materials and measuring directly the
friction of a grain with the wall, we accumulated several
evidences leading us to conclude that the force dependence with
driving velocity and humidity is strongly related to the bead-wall
friction properties: (i) in the steady limit grain/wall tribology
measurements show a friction force increasing with humidity and
velocity, (ii) in the stick-slip regime, the blockage enhancement
can be related to humidity induced aging of the bead-wall
friction, (iii) the hysteretic transition mechanism from
stick-slip to steady-sliding is similar to the one observed in
solid-on-solid experiments \cite{Brockley,Heslot94}.

In spite of a dominant surface effect, we could also identify
contributions of bulk structurations. For example, we related the
transitory part of the response to pushing to a dependance of the
coefficient of redirection between horizontal and vertical stress
with packing fraction. Also a clear difference of the mechanical
parameters extracted from pushing and pulling experiments shows a
contribution from restructuration of about 20\%.

A second system, made of polydisperse assemblies of glass beads,
was investigated. We emphasize the onset of a new complex
dynamical behavior, i.e. the large distribution of blockage forces
evidenced in the stick-slip regime close to the transition.

\section{Acknowledgments}
We acknowledge collaboration with C. Fond, C. Gauthier and E.
Kolb. We thank J. Lanuza and P. Lepert for technical support. We
are grateful to Professor T. Baumberger, Professor R.P. Behringer,
and Professor C. Caroli for many interesting discussions.


\begin{thebibliography}{99}
\bibitem{othermodels}There are other models for
$\theta$, some of them are recessed by Ruina \cite{Ruina}.
\end{thebibliography}
\end{document}